# Diffusion and solvation dynamics of ions in water: beyond the Brownian approximation

Ian C. Bourg,[1,2,*] Minh-Thê Hoang Ngoc,[1] Thomas R. Underwood,[1] Carlos Vega,[3]

[1]Department of Civil and Environmental Engineering, Princeton University, Princeton, New Jersey 08544, USA

[2]High Meadows Environmental Institute, Princeton University, Princeton, New Jersey 08544, USA

[3]Dpto. Química Física I, Fac. Ciencias Químicas, Universidad Complutense de Madrid, 28040 Madrid, Spain

*Corresponding author. Email: bourg@princeton.edu

**Abstract**

The coupled dynamics of ions and water molecules in their first hydration shell impact a variety of processes including ion diffusion, selective ion transport in water-filled nanopores, and the kinetics of ion-pairing, ion adsorption, and metal-ligand binding reactions. In this work, we study these coupled dynamics for alkali metals (Li, Na, K, Rb, Cs), alkaline Earth metals (Mg, Ca, Sr, Ba), and chloride through the lens of their dependence on ion isotopic mass. Results are validated against previous measurements of the isotopic mass-dependence of ion diffusion coefficients in water and previous *ab initio* calculations of ion high-frequency dynamics in water. We find that the vibrational power spectra of ions in water consistently exhibit either two or three peaks, i.e., ions have several rattling frequencies within their solvations shells as previously reported for a subset of the species examined here. These frequencies have different sensititivies to isotopic mass that may serve as signatures of ion solvation processes (such as the tendency of ions to orient their first-shell water molecules) and that also may relate to Hofmeister-like effects including the relative affinity of different metals for ribonucleic acid (RNA).

## 1. Introduction

Ion solvation plays important roles in a wide variety of phenomena in natural, biological, and engineered systems, from protein folding to colloid aggregation, ion-ion separation, planetary geochemistry, contaminant fate and transport, battery technology, and critical element recovery.[1,2] A recurrent question in our fundamental understanding of these phenomena is how solvation impacts ion dynamics including self-diffusion,[3,4] ligand exchange kinetics,[5-8] mineral dissolution and precipitation rates,[9-11] interfacial mass transfer,[12-14] and ion transport through biological and synthetic membranes.[15-17]

Despite extensive previous work on the influence of aqueous solvation on ion dynamics, at least one significant feature of this relation remains largely unexamined: extant theories generally assume that the high-frequency rattling motions of ions within their first hydration shell are entirely decoupled from dynamic properties evaluated in the zero-frequency limit, such as ion self-diffusion coefficients ($D_i$) or the residence times of water molecules in the first hydration shell ($\tau_{\mathrm{w}}$).[18-22] In other words, ions are often theoretically represented as infinitely massive Brownian particles.[23]

Several lines of evidence suggest that the approximation presented above is not entirely appropriate. Theoretical calculations using mode-coupling theory indicate that the $D_i$ values of ions in water should be sensitive to their mass-dependent high-frequency motions.[24] Generalized Langevin dynamics calculations yield the same conclusion in the case of ion pair dissociation kinetics.[8] Experiments and molecular dynamics (MD) simulations have confirmed that ion mass ($m_i$) significantly modulates $D_i$ and $\tau_{\mathrm{w}}$.[4,25-28] The $m_i$-dependence of $D_i$ and $\tau_{\mathrm{w}}$ observed in MD simulation and laboratory experiments, though typically small (on the order of a few per mille), can play an important role in constraining isotope geochemistry reconstuctions of field- and global-scale biogeochemical processes based on subtle variations in isotopic ratios measured in natural systems.[28-33]

One limitation of the experimental and MD simulation studies highlighted above is that they observed the $m_i$-dependence of $D_i$ and $\tau_{\mathrm{w}}$ only in the zero-frequency limit. In other words, they demonstrate that the high-frequency translational modes of ions within their hydration shell modulate static transport properties, but they did not directly examine these modes. Simultaneously, observations of ion solvation dynamics by neutron scattering, terahertz, and correlated vibrational spectroscopy experiments and MD simulations have revealed evidence of ion-water dynamic coupling in aqueous solutions at frequencies on the order of 1 to 10 THz, but they did not evaluate the impacts of these couplings on static transport properties.[34-37] Finally, several studies have reported significant differences in ion-water interactions (friction coefficients, memory functions, $\tau_{\mathrm{w}}$ values) between mobile and immobile ions, but they did not examine a range of ion mobilities as feasible by systematically modulating ion mass.[4,23,38] The present study aims to bridge this observational gap by characterizing how $m_i$ influences both the high-frequency translational dynamics of ions in water and their static transport properties for a wide range of isotopic masses.

Molecular dynamics simulations are well suited to shed light on the high-frequency translational modes of hydrated ions for two reasons. First, they enable a precise, direct evaluation of the velocity autocorrelation functions of aqueous solutes, $C_v(t)$, and the associated power spectra $C_v(\omega)$.[39] In contrast, experiments provide indirect information on $C_v(\omega)$ in comparatively narrower frequency ranges,[36,37] while ab initio MD (AIMD) simulations have higher statistical noise due to their shorter duration.[35] Relative to experimental approaches, atomistic simulations have the additional advantage that they enable manipulating atomic mass over a much wider range than possible using only existing stable isotopes of any given element.[27,28,33,40] The most important downside of classical MD simulations is that their predictions are sensitive to the choice of inter-atomic potential model.[41,42]

In the present study, we build on the work described above by using MD simulations to characterize the static transport properties and high-frequency translational dynamics of ten monoatomic ions at infinite dilution in bulk liquid water: five alkali metals (Li, Na, K, Rb, Cs), four alkaline earth metals (Mg, Ca, Sr, Ba), and one halide (Cl). All ten ions exhibit two or three high-frequency translational modes in the THz domain with different sensitivities to solute isotopic mass. We compare this sensitivity to static and dynamic features of ion solvation including $D_i$, $\tau_{\mathrm{w}}$, the structure of the first hydration shell, and the mean square force $\langle F^2 \rangle$ experienced by ions in water.

## 2. Methods

Molecular dynamics simulations of systems containing 1000 $H_2O$ molecules and one ion were carried out with the code LAMMPS.[43] Simulations used the Verlet algorithm with a time-step of 1 fs and periodic boundary conditions. Water molecules were kept rigid using the SHAKE algorithm.[44] For each ion, a system containing the major isotope of the ion of interest was equilibrated for 1 ps in the NVE ensemble, 9 ps in the NVT ensemble at 298 K, and 490 ps in the NPT ensemble at 1 bar and 298 K using a Nose-Hoover thermostat and barostat. The average volume of the simulation cell was calculated during the final 400 ps of equilibration, and the simulation cell volume was rescaled to this average volume at the end of equilibration. The isotopic mass of the ionic solute was then adjusted to one of five different values ranging from 5 to 140 Da, and each of the five resulting systems was further equilibrated for 200 ps in the NVT ensemble at 298 K. This equilibration procedure resulted in 50 different systems (i.e., ten different solutes, with five isotopes of each solute). Finally, each of the 50 systems was simulated for 75 ns in the NVT ensemble at 298 K.

Interatomic interactions were modeled using the TIP4P2005 water model[45] with the MADRID-2019 model of ionic solutes.[46-48] This set of interatomic potentials uses a scaled charge representation of pairwise Coulomb interaction[49-51] with effective charges of $\pm$0.85 e for monovalent ions and $\pm$1.7 e for divalent ions. These Coulomb interactions are combined with a standard Lennard-Jones (LJ) 6-12 representation of Van der Waals interactions. Pairwise interactions were truncated at a distance of 1.2 nm. Coulomb interactions beyond this cutoff were evaluated using the particle-particle, particle-mesh (PPPM) algorithm with a precision of 99.9%.

Interatomic interaction potentials were assumed to be invariant with isotopic mass in accordance with the Born-Oppenheimer approximation. This approximation is widely used in both classical and ab initio MD simulations of isotope effects,[27,33,52,53] with the notable exception of path-integral MD (PIMD) simulations.[54] It should represent a negligible source of inaccuracy in comparisons with experimental results, as the de Broglie wavelength of all real isotope examined here (with the exception of $^{7}Li^{+}$) is more than 25 times the nearest-neighbor ion-water distance. This approximation enables us to isolate the classical mechanical impacts of $m_i$ on system dynamics.

Ion velocity vectors $\boldsymbol{v}(t)$ were stored every 1 fs for calculation of the velocity autocorrelation function, $C_v(t) = \langle \boldsymbol{v}(0) \cdot \boldsymbol{v}(t) \rangle$, where brackets represent an average over all time intervals of length $t$. The frequency-dependent power spectrum $C_v(\omega)$ was calculated as the real part of the Fourier transform of $C_v(t)$. Spectral leakage was reduced by applying a Blackman window with $t_{\max}$ = 4 ps prior to Fourier transform calculation.

Almost all calculated power spectra exhibited two or three peaks. The only exceptions were the spectra of four of the lightest divalent ions ($^{6}$Mg, $^{12}$Mg, $^{5}$Ca, and $^{5}$Sr), which exhibited a fourth peak or shoulder with a maximum frequency of ≈ 28 THz. We hypothesize that this fourth mode reflects coupling with water's librational modes, which drop off sharply at frequencies above 28 THz. Since this mode was observed only for four of the 50 solutes examined here, and only for isotopes that do not exist in reality, it was not examined in this work.

Power spectra were fitted as the sum of either two or three peaks through linear least-squares minimization. The fitting procedure used a cutoff frequency 2 THz higher than that of the third (i.e., highest-frequency) mode. For $^{6}$Mg, $^{12}$Mg, and $^{5}$Sr, the cutoff was set instead to the peak frequency of the third mode to avoid the fourth mode described above. For $^{5}$Ca, the third mode presents as a shoulder on the fourth mode peak and its frequency is highly uncertain. All spectra were fitted using a combination of two or three Lorentzian functions (i.e., damped harmonic oscillations), $I(\omega) = A/\{1 + [(\omega - \omega_0)/w]^2\}$, where $\omega_0$, $A$, and $w$ are the peak position, amplitude, and width. In cases where three modes could not be unambiguously distinguished (i.e., for K, Rb, Cs, and Cl with isotopic masses greater than 30 Da), the fitting procedure used only two modes. Since the second mode almost invariably had the smallest amplitude in the three-mode fitting results, the two-mode fitting results were interpreted as indicating the frequencies of the first and third modes.

The zero-time limit of $C_v(t)$ was used to calculate the mean-square translational force on the solute, $\langle F^2 \rangle$, through a second-order Taylor expansion of the normalized velocity autocorrelation function: $C_v(t)/C_v(0) = 1 - (t^2/2) \times a$, where $a = \langle F^2 \rangle / 3 m_i k_{\mathrm{B}} T$. Results relate to the translational Einstein frequency $\omega_{\mathrm{E}} = \left( \int_0^\infty \omega^2 C_v^*(\omega) d\omega \right)^{0.5} = a^{0.5}/2\pi$, where $C_v^*(\omega)$ is the Fourier transform of $C_v(t)/C_v(0)$, and to the quantum mechanical zero-point energy (ZPE) of the solute, which determines the isotopic mass dependence of the ion's chemical potential.[40,55] In the case of multi-atomic molecules, accurate prediction of $\langle F^2 \rangle$ (or, equivalently, $\omega_{\mathrm{E}}$) requires quantum mechanical calculations to correctly evaluate the frequency of intramolecular vibrational modes.[56] For monoatomic solutes, however, results obtained for noble gases in water indicate that classical MD simulations accurately predict $\langle F^2 \rangle$ for Ne, Ar, Kr, Xe.[40]

The time-integral of $C_v(t)$ was used to calculate ion self-diffusion coefficients according to the Green-Kubo relation $D_{i,\mathrm{GK}} = \frac{1}{3} \lim_{\tau \to \infty} \int_0^\tau C_v(t) dt$.[57] The infinite-time limit was approximated by calculating the average value of $D_i$ for $\tau = 3$ to 3.5 ps.

The positions of all atoms were stored every 1 ps for calculation of configurational properties, hydration dynamics, and an alternate calculation of ion self-diffusion coefficients. Radial distribution functions were calculated between the ion and water O atoms ($g_{i\mathrm{O}}(r)$). The first maximum ($r_{\mathrm{max}}$) and minimum ($r_{\mathrm{min}}$) of $g_{i\mathrm{O}}(r)$ were used as measures of the size of the first ionic hydration shell. Integration of $g_{i\mathrm{O}}(r)$ from $r = 0$ to $r_{\mathrm{min}}$ was used to calculate the first-shell coordination number $N_{\mathrm{O}}$ of each ion.[57]

In addition to the calculation based on $C_v(t)$, ion self-diffusion coefficients also were determined using the Einstein relation $D_{i,\mathrm{E}} = \frac{1}{6} \lim_{\Delta t \to \infty} \frac{\langle \Delta l^2 \rangle}{\Delta t}$, where $\langle \Delta l^2 \rangle$ is the mean-square displacement of the ion over a time interval $\Delta t$. The infinite-time limit was approximated by calculating the slope of $\langle \Delta l^2 \rangle$ vs. $\Delta t$ from $\Delta t = 2$ to 10 ps.[58] Diffusion coefficients $D_{i,\mathrm{E}}$ and $D_{i,\mathrm{GK}}$ were mutually consistent and exhibited similar statistical error.

Direct observations of $D_i$ by MD simulation exhibit a known artefact associated with the periodic boundary conditions of the simulated systems. Results obtained in this work were corrected for this artefact using the theoretical expression $D_i = D_{i,\mathrm{GK}} + k_\mathrm{B}T\xi/6\pi\eta L$.[59] In this equation, $k_\mathrm{B} = 1.380658 \times 10^{-23}$ J K$^{-1}$ is Boltzmann's constant, $T = 298$ K is absolute temperature, $\xi \approx 2.837297$ is a constant associated with the lattice sum calculation,[60] $\eta = 0.855 \times 10^{-3}$ Pa s is the viscosity of TIP4P/2005 water,[61] and $L$ = 31.136 Å is the average simulation cell size in this study (with a standard deviation of 0.008 Å between systems with different ions). In the conditions of this work, the correction factor $k_BT\xi/6\pi\eta L$ had a value of $0.23 \times 10^{-9}$ m$^2$ s$^{-1}$ with error associated predominantly with the value of $\xi$, which we estimate conservatively at roughly ±10%.

The rate constant for water exchange between the first and second hydration shells, $k_\mathrm{w}$, was determined by calculating the autocorrelation function $C_\mathrm{w}(t)$ that characterizes the likelihood that a water molecule located in the first solvation shell at any given time (defined by an ion-water O distance shorter than $r_\mathrm{min}$) remains continuously located in the first hydration shell during a following timespan of length $t$. Although more rigorous treatments of barrier recrossing events have been developed,[62] for simplicity short excursions outside the first shell for less than 2 ps were neglected as in many previous studies.[20,63,64] Results were fitted through linear regression to a first-order decay relation, $\ln C_\mathrm{w}(t) = -k_\mathrm{w}t + \ln C_{\mathrm{w},0}$ and converted to a residence time of water molecules in the first shell, $\tau_\mathrm{w} = 1/k_\mathrm{w}$.

Confidence intervals on $D_i$ and $k_\mathrm{w}$ were calculated by dividing each 75 ns simulation into three 25 ns blocks.[57] Statistical uncertainties are reported as two standard errors.

## 3. Results

### *3.1. Coordination structure of ions in water*

Results on the first-shell coordination structure of ions in water are reported in Table 1 (for $r_\mathrm{max}$ and $N_\mathrm{O}$) and in Supplementary Information Table S1 (for $r_\mathrm{min}$). Results are broadly consistent with values reported based on neutron diffraction with isotope substitution (NDIS) or extended X-ray absorption fine-structure (EXAFS) spectroscopy characterization of aqueous solutions at salinity ≤ 1 M (Table 1). For most ions examined here, the MADRID-2019 model accurately predicts $N_\mathrm{O}$. Simulation results slightly under-predict experimental $r_\mathrm{max}$ values (by 0.07±0.06 Å on average), possibly because of the excessive stiffness of the repulsive part of the LJ 6-12 potential. Conversely, previous studies suggest that AIMD methods tend to overestimate experimental $r_\mathrm{max}$ values by 0.1±0.1 Å on average,[71,101] such that our simulations match more closely with experiments than with AIMD results. The identical values of $r_\mathrm{min}$ and $r_\mathrm{max}$ for K and Rb, while somewhat unexpected, are consistent with AIMD calculations indicating that both ions exhibit an inner coordination shell with six water molecules and that the size of this inner shell is identical for K and Rb, even as the $r_{\max}$ values predicted by AIMD differ by 0.2 Å.[82]

**Table 1. Solvation structure of ions in water.** Results from this work have a precision of ±0.01 Å and 0.1 for $r_{\max}$ and $N_O$, respectively, based on comparison with previous results obtained with the MADRID-2019 model.[46-48] Results from previous studies are restricted to EXAFS or NDIS experiments at salinity ≤ 1 M, AIMD simulations of individual ions at infinite dilution, and classical MD calculations carried out using the highly accurate MB-nrg multi-body potential energy model. All results were obtained at or near standard temperature and pressure.

| | $r_{\max}$ (Å) | | | | $N_O$ | | | |
|---|---|---|---|---|---|---|---|---|
| | This work | Experiments | AIMD | MB-nrg | This work | Experiments | AIMD | MB-nrg |
| Li | 1.84 | 1.96(2)[65], 1.99(3)[66] | 1.948[67], 1.96[35] | 1.93[68] | 4.0 | 4.8(3)[65], 6.0(2)[66] | 3.995[67], 4[69], 4.0[35] | 4.1[68] |
| Na | 2.33 | 2.34(14)[70], 2.37(2)[71] | 2.48(0.12)[71], 2.35[72] | 2.35[73], 2.38[68] | 5.5 | 5.3(0.8)[70], 5.4(1.3)[71] | 5.7(8)[71], 4.6(6)[72] | 5.8[73], 5.8[68] |
| K | 2.74 | 2.65(18)[70], 2.73(2)[74] | 2.82[75], 2.80(5)[76], 2.83[77], 2.73[77] | 2.7[73], 2.73[68] | 6.8 | 6.0(1.2)[70], 6.1(1.0)[74] | 5.7[75], 5.9(2)[76], 6.6[77], 5.9[77] | 6.7[73], 6.7[68] |
| Rb | 2.74 | 2.93(3)[78], 2.98(4)[79], 2.83(1)[80], 2.886(9)[81] | 3.0[82] | 2.88[68] | 6.4 | 5.6(7)[78], 8.0(5)[79], 6[80], 6.5(4)[81] | 6.8(2)[82] | 7.6[68] |
| Cs | 2.86 | 2.98[83], 3.1[84], 3.15(5)[85] | | 3.13[68] | 6.9 | 8.0[83], 8.20(4)[84], 6.73[85] | | 8.7[68] |
| Mg | 1.92 | 2.10(3)[86] | 2.13(6)[87], 2.10[88] | | 6.0 | ~6[86] | 6.01(9)[88] | |
| Ca | 2.39 | 2.43(3)[89], 2.46[86] | 2.37[88], 2.51(7)[90] | | 7.4 | 6.8(1.0)[89], ~8[86] | 5.83(10)[88], 6.67(24)[90] | |
| Sr | 2.60 | 2.57[91], 2.62(3)[92], 2.63(3)[93], 2.63(2)[94] | 2.63[95] | | 8.1 | 7.3(5)[92], 7.7(5)[93], 7.8[91], 8[94] | 8.0[95] | |
| Ba | 2.84 | 2.780(3)[96], 2.81(3)[94] | 2.8[97] | | 9.0 | 7.8(3)[96], 8[94] | 7.8[97] | |
| Cl | 3.03 | 3.11(3)[89], 3.15(15)[70], 3.174(29)[81], 3.225(30)[86] | 3.1[98], 3.24[99] | 3.18[100] | 5.9 | 6.4(1.0)[89], 6.8(3)[84], 6.95(1.10)[70], 7.9(1.1)[81] | 5.8[98] | 8.2[100] |

### *3.2. Self-diffusion coefficients of ions in water*

The self-diffusion coefficients $D_i$ of the major isotopes of each ion are presented in Fig. 1a and in Supplementary Information Table S2. For systems such as the one simulated here that violate electroneutrality, our treatment of long-range Coulomb interactions implies the existence of a neutralizing background with conducting boundary conditions. Previous simulations by Blazquez et al.[47] that used the MADRID-2019 model to simulate individual ions in water yielded statistically identical results as shown in Table S2. Our previous work also verified that identical results are obtained for single Na or Cl ions and for an extremely dilute (by MD simulation standards) charge-neutral system containing 0.01 M NaCl. Results of this work demonstrate good agreement with experimental values compiled by Li and Gregory,[102] with a mean absolute error of 6.7%. The largest discrepancy is observed for Cl, where our simulations underestimate $D_i$ by 17%.

Previous studies have hypothesized[25] and later confirmed[26,27] that the isotopic mass-dependence of $D_i$ follows a power-law relation, $D_i \propto m_i^{-\beta}$, where the inverse power-law exponent $\beta$ has values between 0 and 0.05 for ions and up to 0.2 for noble gases.[27,103] A plot of $\log D_i$ vs. $\log m_i$ for the five isotopes of each ion examined in this work confirms this expectation (Fig. 1b). Values of the inverse power-law exponent $\beta$ obtained by linear regression of $\log D_i$ vs. $\log m_i$ are compared with previous experimental results[25,27,104-106] in Fig. 1c and in Supplementary Information Table S3. Simulation predictions are consistent with measured $\beta$ values for most species except Li and Cl, for which the predictions fall slightly outside the experimental uncertainty range.

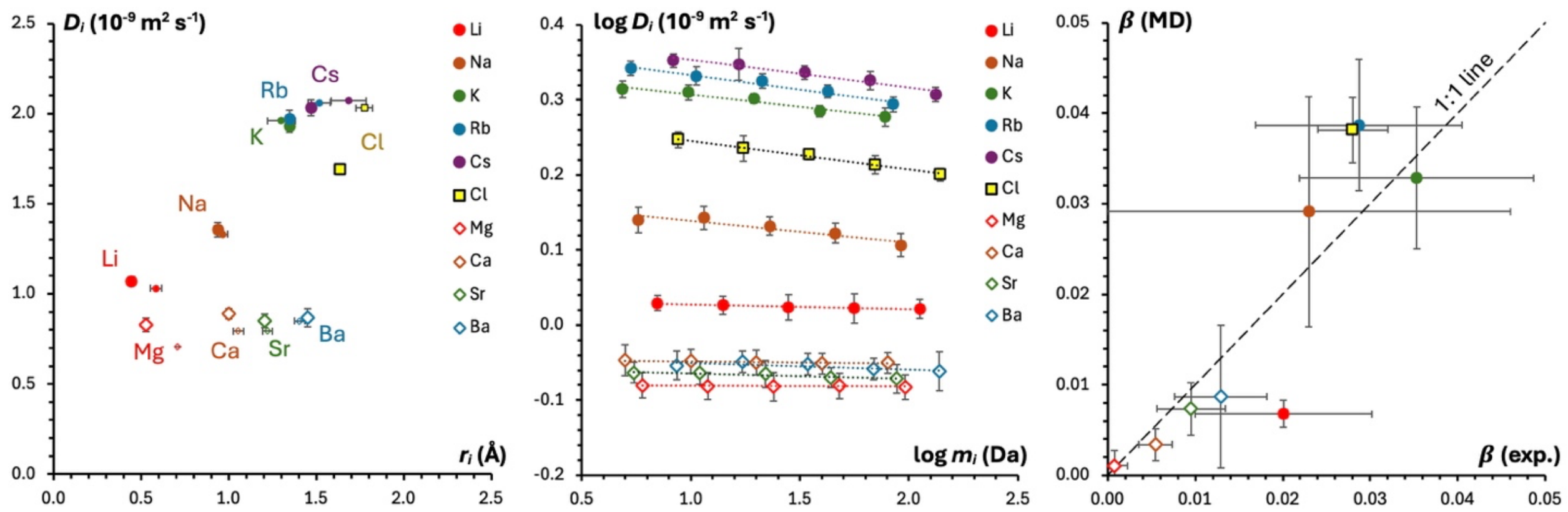

**Figure 1. Ion diffusion in water. (a)** Self-diffusion coefficient $D_i$ ($10^{-9}$ m$^2$ s$^{-1}$) as a function of bare ion radius $r_i$ (Å). Bare radii were calculated as $r_{\text{max}}$ minus the radius of a water O atom (1.39 Å), estimated as half of the peak O-O distance in water in our MD simulations. Large symbols show MD simulation predictions obtained in this work, where $D_i$ values are shown for the most abundant isotope of each ion. Small symbols are based on experimental results (Table 1, Table S2). **(b)** MD simulation predictions of $\log D_i$ as a function of $\log m_i$ for all ions examined in this work. Error bars show 95% confidence intervals. Dashed lines show linear regressions. For all solutes except Mg, results are statistically consistent with an inverse power-law relation, $D_i \propto m_i^{-\beta}$, with $\beta > 0$. **(c)** Comparison of predicted and measured $\beta$-values. The vertical axis shows $\beta$ values determined in this work. The horizontal axis shows the average of experimental values reported in previous studies (Table S3). Where several experimental values exist, horizontal error bars show 95% confidence intervals based on the standard deviation of previous results. Where only one high-precision experimental value exists (Rb, Sr, Ba), error bars were estimated based on a relative standard deviation of 20% inferred from the pairs of high-precision results reported for Li, K, and Ca. Where only one low-precision experimental value exists (Na), we used the 95% confidence interval reported in previous work.

### *3.3. Water ligand exchange kinetics*

The residence times $\tau_{\text{w}}$ of water molecules in the first hydration shell of the major isotope of each ion are presented in Fig. 2a and in Supplementary Information Table S4. Results are broadly consistent with the trends reported in previous MD simulation studies.[20,21,63,64] Major quantitative differences include a significantly greater stability of the first solvation shell around Li, and smaller stability around divalent metals, than reported in most previous MD simulation studies. Comparison with previous studies is complicated by systematic uncertainties associated with the identification of water-exchange event from simulation trajectories[62] and with the indirect evaluation of $\tau_{\mathrm{w}}$ by experimental approaches. Nevertheless, results obtained in this work are generally consistent with AIMD simulations,[88,90,98,107,108] pump-probe spectroscopy measurements,[109] and MD simulations carried out with the MD-nrg model[68,100] as well as with MD simulations of seawater carried out with the MADRID-2019 model,[42] as shown in Table S4.

To the best of our knowledge, a single previous MD simulation study has examined the isotopic mass-dependence of $\tau_{\mathrm{w}}$ in bulk liquid water. Results showed that $\tau_{\mathrm{w}}$ has a power-law dependence on isotopic mass, $\tau_{\mathrm{w}} \propto m_i^{\gamma}$ or, equivalently, $k_{\mathrm{w}} \propto m_i^{-\gamma}$, with $\gamma = 0.05 \pm 0.01$; predicted $\gamma$ values exhibited no significant dependence on temperature (278, 298, or 323 K), water model (SPC/E or TIP4P), or ion type (Li, K, Rb, Ca, Sr, or Ba).[28] Results on $\log \tau_{\mathrm{w}}$ vs. $\log m_i$ obtained in this work broadly confirm the findings of Hofmann et al.[28] for a greater variety of ions (Fig. 2b). However, the greater precision of $\gamma$ values obtained in this work enables new conclusions regarding the sensitivity of $\gamma$ to ion type (Fig. 2c). Specifically, most monovalent ions (Na, K, Cs, Cl) have $\gamma = 0.05 \pm 0.01$, shown as a horizontal dashed line in Fig. 2c, but Li has $\gamma = 0.20 \pm 0.05$ and divalent ions (Ca, Sr, Ba) have $\gamma = 0.03 \pm 0.01$. For Mg, too few water-exchange events were observed to precisely quantify $\gamma$. Results shown in Fig. 2 are presented in tabular form in Tables S1 and S5.

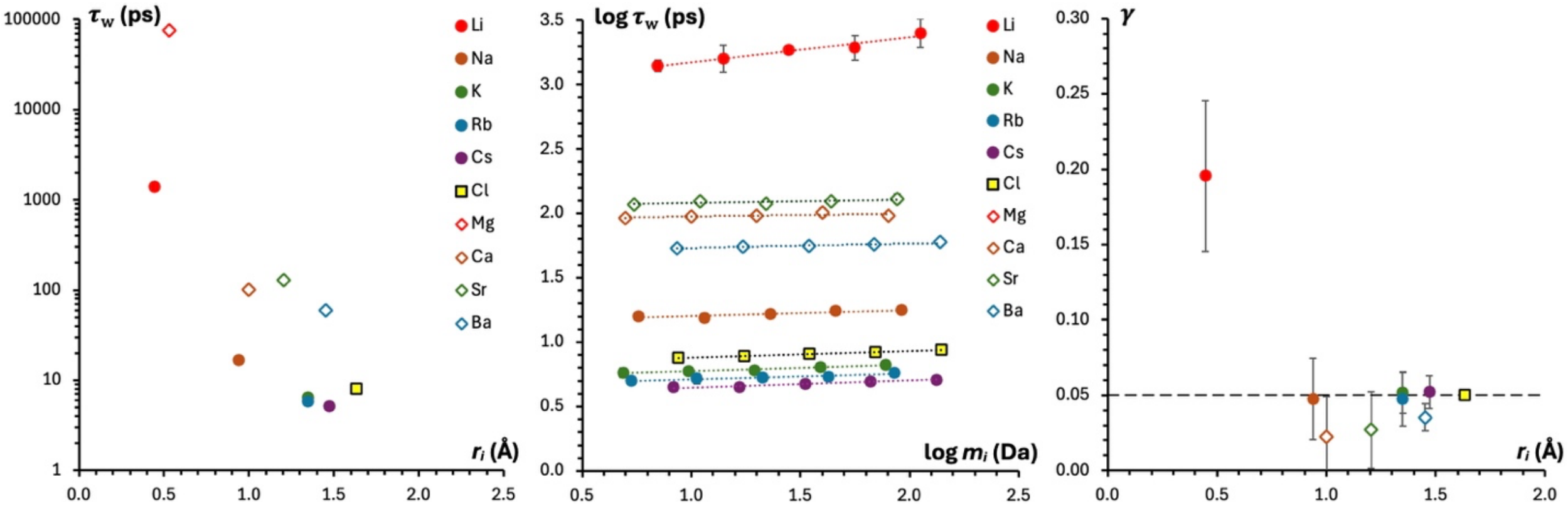

**Figure 2. Residence time $\tau_{\mathrm{w}}$ (ps) of water molecules in the first hydration shell of ions. (a)** $\tau_{\mathrm{w}}$ values of the most abundant isotope of each ion plotted on a logarithmic scale as a function of bare ion radius $r_i$ (Å). All symbols show MD simulation predictions from this work. **(b)** MD simulation predictions of $\log \tau_{\mathrm{w}}$ as a function of $\log m_i$ for all ions examined in this work. Where not visible, vertical error bars are smaller than the symbol size. Dashed lines show linear regressions. For all solutes except Mg (off-scale with $\log \tau_{\mathrm{w}} > 3.5$), results are statistically consistent with a power-law relation with exponent $\gamma > 0$. **(c)** Predicted $\gamma$-values plotted as a function of bare ion radius $r_i$ (Å). The horizontal dashed line indicates the average value of $\gamma = 0.05 \pm 0.01$ determined in previous work for six ions (Li, K, Rb, Ca, Sr, and Ba).[28]

### *3.4. Velocity autocorrelation functions*

Representative results on the $m_i$ dependence of $C_v(t)$ are shown in Fig. 3a in the case of five real or hypothetical isotopes of Li. As expected, $C_v(t)$ exhibits a strong dependence on ion mass. The value at time zero, $C_v(t = 0) = \langle \boldsymbol{v}^2 \rangle$, is inversely proportional to $m_i$ as expected from the equipartition principle [$m_i \langle \boldsymbol{v}^2 \rangle = 3k_\mathrm{B}T$]. The duration of the ballistic regime—the initial plateau in $C_v(t)$ vs. log($t$)—increases with isotopic mass. Light isotopes exhibit significant recoil as shown by the short-time sign-reversal of $C_v(t)$, whereas heavy isotopes do not.

Figure 3b shows one-third of the cumulative integral of $C_v(t)$, which yields $D_{i,\mathrm{GK}}$ at large times. Results highlight how the large influence of $m_i$ on ion dynamics on time-scales < 0.3 ps becomes *almost* insignificant on time-scales > 3 ps, as expected if the high-frequency rattling motions of ions within their solvation shell are fully decoupled from their zero-frequency diffusive motion. Four distinct time-scales are apparent in Fig. 3b: ballistic motion at $t < 0.01$ ps; rattling motions at $t < 0.3$ ps; a significant 'backflow' at $t = 0.3$ to 3 ps; and a long-time diffusive regime at $t > 3$ ps.

Figure 3c shows the vibrational power spectra $C_v(\omega)$ obtained from the $C_v(t)$ curves in Fig. 3a. Lithium ions exhibit three translational modes with frequencies of ~1, 5, and 15 THz, referred to hereafter as the $\omega_1$, $\omega_2$, and $\omega_3$ modes. These three modes can be discerned in the $^7$Li power spectrum reported by Lyubartsev et al.[35] based on short (20 ps) AIMD simulations, where they appear at roughly 1.5, 5, and 12 THz. Two or three analogous peaks were reported based on MD simulations of ions in water by Funkner et al.[36] (for $^{24}$Mg, $^{40}$Ca, $^{88}$Sr, and $^{35}$Cl) and by Lammers et al.[33] (for $^{40}$Ca and a hypothetical $^{10}$Ca isotope).

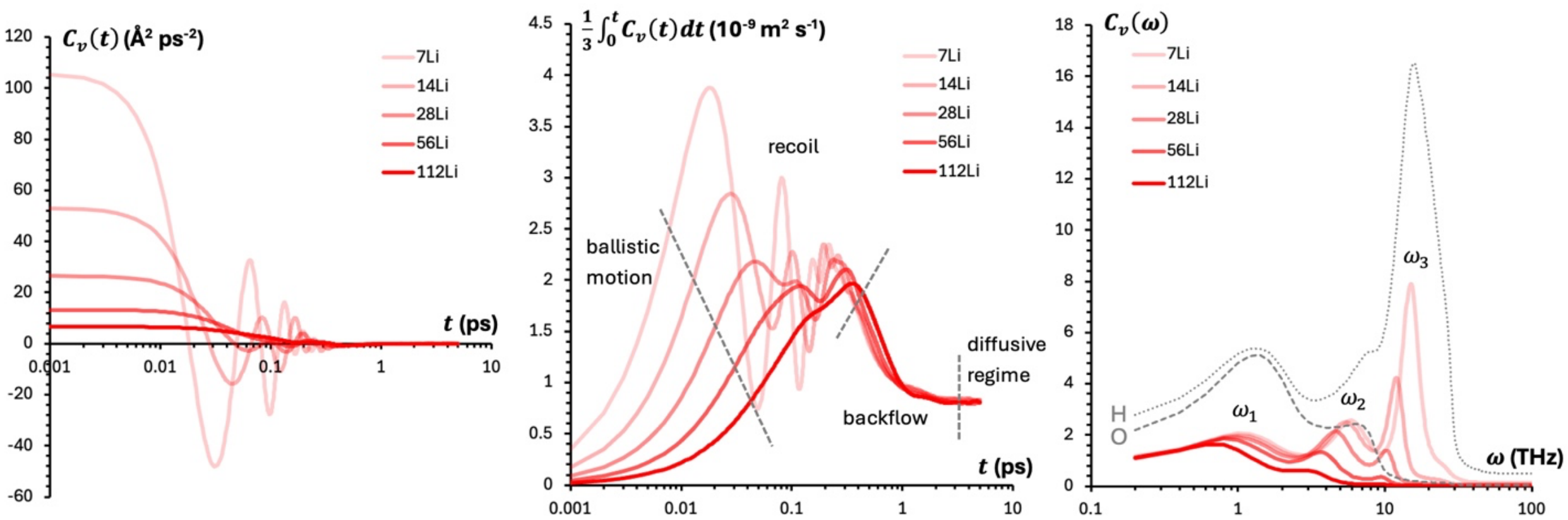

**Figure 3. Velocity autocorrelation function $C_v(t)$ and associated properties in the case of five real or hypothetical isotopes of Li⁺. (a)** $C_v(t)$ plotted as a function of time, with a logarithmic time scale. **(b)** Cumulative integral of $\frac{1}{3}C_v(t)$ as a function of time, with a logarithmic time scale. **(c)** Power spectrum $C_v(\omega)$ plotted as a function of frequency $\omega$ (THz) with a logarithmic frequency scale. Lithium exhibits three translational modes labeled $\omega_1$, $\omega_2$, and $\omega_3$. The gray dashed and dotted lines show the power spectra of water O and H atoms in pure water.

In Fig. 3c, we also report the power spectra of water O and H atoms obtained using the TIP4P/2005 water model as dashed and dotted gray lines. The water O spectrum exhibits a main peak at 1.31±0.04 THz and a broad shoulder at 4 to 7 THz associated with bending and stretching modes of the hydrogen bond (H-bond) network of liquid water, respectively.[37,110] The water H spectrum also has a prominent peak at 15.80±0.02 THz associated with librational motions. Water O and H power spectra calculated in this work are consistent with previous experimental and AIMD observations[35,37,111,112] and with classical MD simulation results obtained with other water models.[39,113,114]

### *3.5. Mean square force*

Predictions of the mean square force $\langle F^2 \rangle$ experienced by each ion are shown in Fig. 4a. The magnitude and trends are consistent with previous observations.[115] Results generally indicate a tighter hydration shell around smaller, more highly charged ions. One exception is observed for Mg, which experiences a lower $\langle F^2 \rangle$ value than Ca, possibly indicating that the tightness of its solvation shell is limited by steric repulsion between first-shell water molecules. The other exception is observed for Cl, which experiences a higher $\langle F^2 \rangle$ than Cs, perhaps because of the unequal distance between water positive and negative partial charge sites and the Van der Waals center of water (i.e., Cl can more closely approach water H atoms than Cs can approach water O atoms). When plotted as a function of $|z_i|/r_{\max}$ (where $|z_i|$ is the absolute value of the ionic charge in elementary charge units) as an approximate measure of the electrostatic field strength imposed by the ion at the distance of a first-shell water O atom, the results collapse into a nearly linear relationship (Fig. 4b).

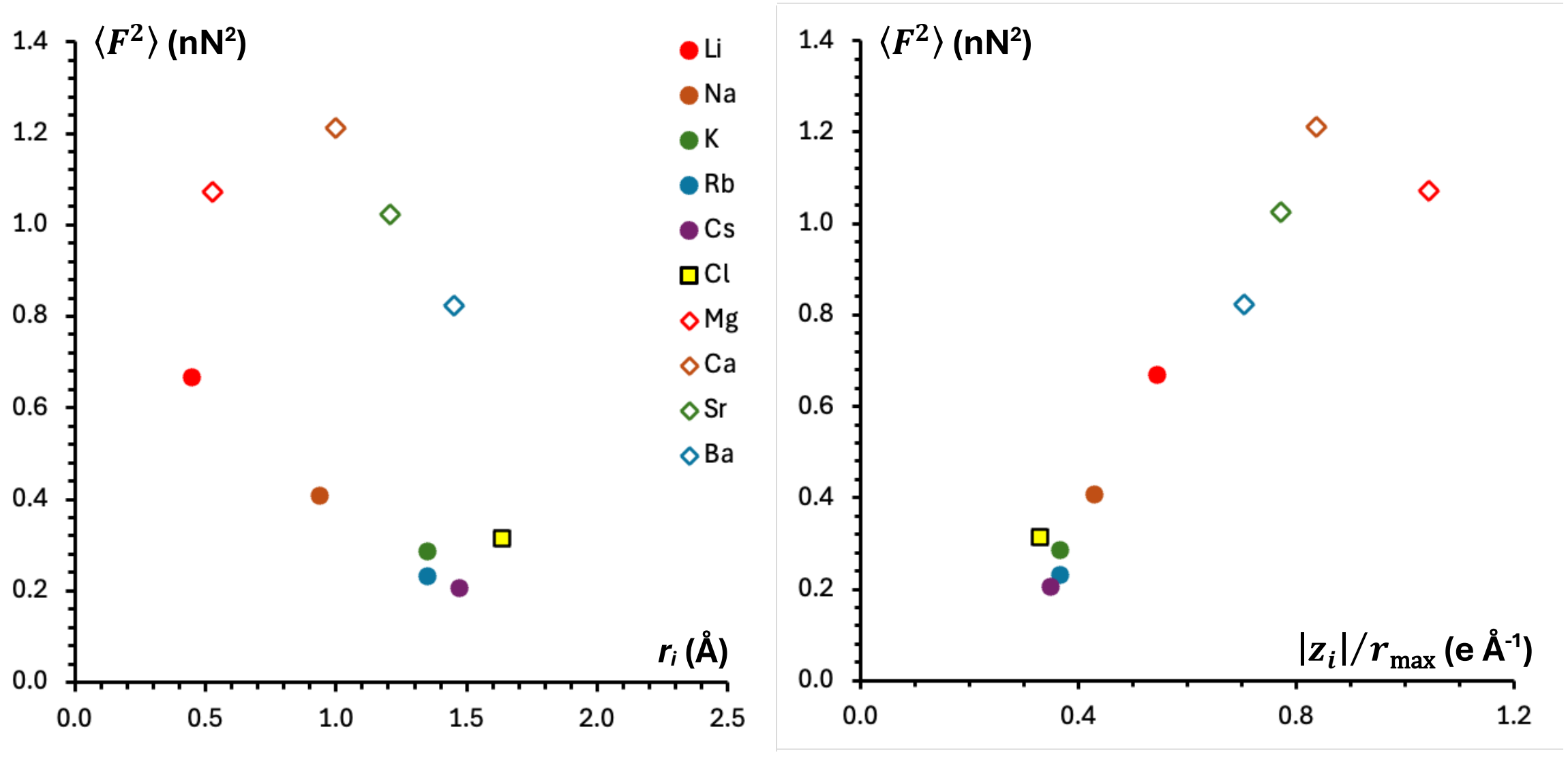

**Figure 4. Mean square force experienced by the ions examined in this work. (a)** $\langle F^2 \rangle$ plotted as a function of bare ion radius $r_i$ (Å). **(b)** $\langle F^2 \rangle$ plotted as a function of $|z_i|/r_{\max}$ (e Å$^{-1}$).

Results shown in Fig. 4 enable calculation of the equilibrium isotope fractionation associated with transferring each ion from a vacuum reference state into pure liquid water at 298 K.[40,55] Results are presented in Table 2 as $\delta_\mathrm{w}$ values calculated for the two most abundant stable isotopes of each ion, i.e., the expected fractional enrichment of the heavy isotope in the aqueous phase relative to the vacuum reference state, conventionally expressed in ‰ units.[101] The fractional enrichment was calculated as $\delta_\mathrm{w} = \frac{\hbar^2}{24(k_\mathrm{B}T)^3}\left(\frac{1}{m_\mathrm{l}} - \frac{1}{m_\mathrm{h}}\right)\langle F^2\rangle$, where $\hbar$ = 1.05457266 × 10$^{-34}$ J s is the reduced Planck constant and $m_\mathrm{l}$ and $m_\mathrm{h}$ are the mass of the light and heavy isotope.

Results presented in Table 2 indicate that classical MD simulations either accurately predict of slightly underestimate ab initio predictions of $\delta_\mathrm{w}$ (and, equivalently, $\langle F^2\rangle$) for most ions except Mg. More precisely, predicted $\delta_\mathrm{w}$ values are smaller than those reported based on ab initio calculations by 5 to 10% for monovalent ions (Li, K), 20% for most divalent cations (Ca, Sr), and 40% for Mg. Ab initio studies that evaluated different methodological approaches (such as different basis sets or exchange-correlation functionals) reported systematic errors on the order of ±25% (Rustad et al., 2010; Colla et al., 2018). Ab initio studies that used the greatest number of explicit water molecules yielded the closest agreement with our classical MD simulation predictions.[75,116]

**Table 2. Equilibrium isotope fractionation factor $\delta_\mathrm{w}$ (‰) calculated for the two most abundant stable isotopes of each ion examined in this work upon partitioning between vacuum and pure liquid water at 298 K.** Results are not reported for ions that have only one stable isotope (Na, Cs). Ab initio predictions were obtained using Hartree-Fock (HF), Møller-Plesset perturbation (MP2), or density functional theory (DFT) calculations using either periodically-replicated simulation cells containing 63 or 64 explicit water molecules[75,116] or clusters with one or two explicit hydration shells in a continuum solvent.[87,101,117-122]

| | This work | Ab initio |
|---|---|---|
| $^{7}$Li/$^{6}$Li | 63.5 | 66[117], 68[118] |
| $^{41}$K/$^{39}$K | 1.43 | 1.61[116], 1.56(21)[75] |
| $^{87}$Rb/$^{85}$Rb | 0.25 | |
| $^{26}$Mg/$^{24}$Mg | 13.8 | 22(2)[101], 22.8(2)[87], 24.2(2.3)[119] |
| $^{44}$Ca/$^{40}$Ca | 11.0 | 12.7[120], 15.1[119] |
| $^{88}$Sr/$^{86}$Sr | 1.08 | 1.36[121], 1.39(4)[122] |
| $^{138}$Ba/$^{137}$Ba | 0.17 | |
| $^{37}$Cl/$^{35}$Cl | 1.94 | |

### *3.6. Ion power spectra*

In most of the range of isotopic mass examined in this study, all ions exhibited either three (Li, Na, Mg, Ca, Sr, Ba) or two rattling modes (K, Rb, Cs, Cl). All $C_v(\omega)$ spectra obtained in this work are shown in Fig. 5. Results are consistent with previous MD simulation observations that the power spectra of Ca, Sr, and Ba exhibit three frequency modes.[33,36] Results also are consistent with time-domain polarization-resolved coherent Raman scattering observations by Heisler and Meech,[123] which showed a vibrational mode at 5.2 THz—ascribed to fluctuations in the OH···Cl⁻ hydrogen bond length—in a 1 M NaCl solution. For comparison, our simulations yield $\omega_3 = 5.08$ THz for $^{35}$Cl.

As noted in the Methods section, four of the 50 solutes ($^{6}$Mg, $^{12}$Mg, $^{5}$Ca, and $^{5}$Sr) presented an additional, higher-frequency peak or shoulder that is not examined in this work. As $m_i$ approaches 100 Da, the high-frequency $\omega_2$ and $\omega_3$ modes are strongly attenuated and the power spectra become dominated by the $\omega_1$ mode. This transition coincides with the disappearance of oscillations in $C_v(t)$ as illustrated in Fig. 3a in the case of the heaviest Li isotope examined in this work.

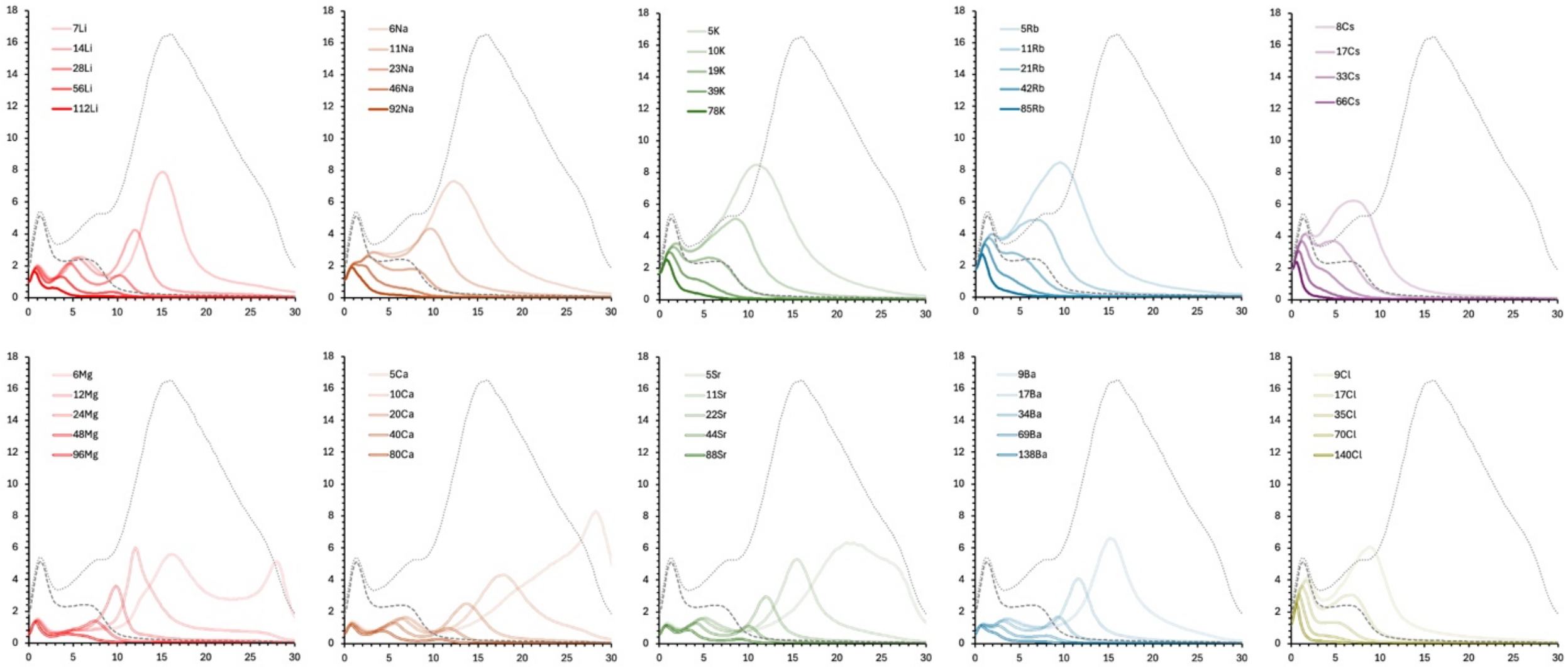

**Figure 5. Power spectra of all ions examined in this work as a function of frequency (THz). (a)** $Li^+$. **(b)** $Na^+$. **(c)** $K^+$. **(d)** $Rb^+$. **(e)** $Cs^+$. **(f)** $Mg^{2+}$. **(g)** $Ca^{2+}$. **(h)** $Sr^{2+}$. **(i)** $Ba^{2+}$ **(j)** $Cl^-$. The power spectra of water O and H atoms are shown as gray dashed and dotted lines on each plot.

Values of the $\omega_1$, $\omega_2$, and $\omega_3$ frequencies of different ions are reported as a function of $m_i$ in Fig. 6 and Table S5. The frequency of the $\omega_1$ mode, shown in Fig. 6a, is close to the main translational mode of water O atoms at 1.31±0.04 THz, suggesting that it corresponds to the coupling of ion motions to the water H-bond 'bend' mode, a collective mode that involves water molecules separated by up to ≈ 6 Å, i.e., second-nearest neighbors.[110] This $\omega_1$ mode exhibits a clear divergence between two groups of solutes. For large monovalent ions (K, Rb, Cs, Cl), the frequency of the $\omega_1$ mode approaches 1.7±0.1 THz at low $m_i$ and equals the frequency of the water H-bond bend mode when $m_i$ equals the mass of a water molecule, suggesting that the ion participates in water's collective H-bond bend mode as if it were a water molecule, albeit with a different mass. For the other ions examined here (Li, Na, Mg, Ca, Sr, Ba), the frequency of the $\omega_1$ mode remains below that of the water H-bond bend mode and approaches 1.0±0.1 THz at low $m_i$, suggesting that these ions have a weaker coupling to the water's collective H-bond bend mode. This conclusion is further supported by the amplitude of the $\omega_1$ mode (Fig. 5), which is significantly greater and more sensititve to $m_i$ for ions in the first group identified above (i.e., K, Rb, Cs, Cl).

The frequency of the $\omega_2$ mode is presented in Fig. 6b for the five ions that exhibit a clearly defined intermediate frequency mode (Li, Na, Ca, Sr, Ba). These ions include every ion in the second group identified above (i.e., ions with a low $\omega_1$ frequency) with the exception of Mg, for which a well-defined $\omega_2$ peak is not visible even though the intensity of the $C_v(\omega)$ curves near 5 THz suggest that the $\omega_2$ mode is present (Fig. 5f). This observation suggests that these ions, though weakly coupled to water's collective H-bond bend mode, are strongly coupled to water's pairwise H-bond 'stretch' mode. All five ions in this group follow a similar relation between $\omega_2$ and $m_i$, with a frequency that increases in the order Na < Ba < Sr ~ Li < Ca.

Finally, the frequency of the $\omega_3$ mode is presented in Fig. 6c as a function of $m_i$ on a double logarithmic plot. Kinetic-theory-like relations of the form $\omega_3 \propto m_i^{0.5}$ and $\omega_3 \propto \mu_i^{0.5}$ (where $\mu_i = \frac{m_i m_w}{m_i + m_w}$ is the reduced mass of the ion and one water molecule) are shown as dashed and dotted lines. Although some ions examined here have $\omega_3$ values consistent with an inverse square-root dependence on $m_i$ (such as Cs), others show a relation consistent with an inverse square-root dependence on $\mu_i$ (such as Li).

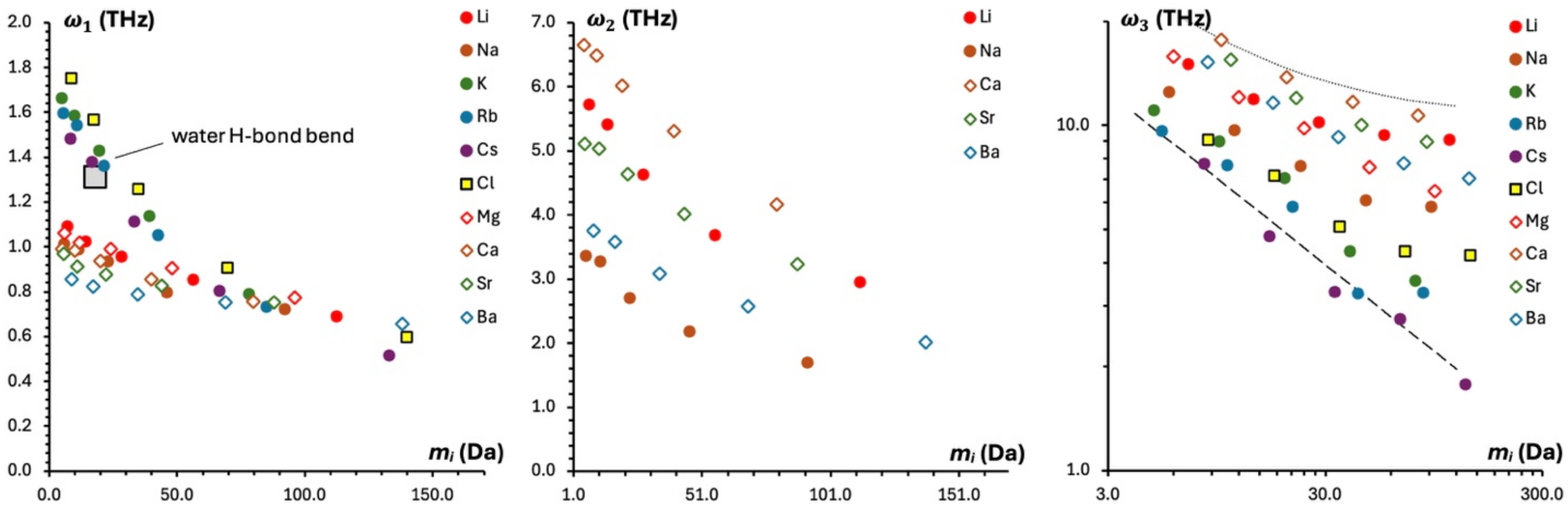

**Figure 6. Mass dependence of the $\omega_1$, $\omega_2$, and $\omega_3$ frequencies (THz) of different ions. (a)** $\omega_1$ as a function of $m_i$. **(b)** $\omega_2$ as a function of $m_i$ for the five ions that exhibit a clear $\omega_2$ peak. **(c)** $\omega_3$ as a function of $m_i$ on a double logarithmic scale. The dashed and dotted lines show kinetic-theory-like relations $\omega_3 \propto m_i^{0.5}$ and $\omega_3 \propto \mu_i^{0.5}$.

## 4. Discussion

Our results confirm that hydrated ions exhibit a significant dependence on $m_i$ of both their high-frequency rattling modes (and hence their ZPE) and of their zero-frequency dynamic properties (self-diffusion coefficient and water ligand exchange kinetics). Despite the systematic examination presented in this work, a relationship between the mass dependence of high-frequency and zero-frequency dynamics does not clearly emerge.

An original result reported in this work is the finding of a sharp divergence between the $\omega_1$ frequencies of two groups of ions in Fig. 6a, where low $\omega_1$ values are observed for Ba, Sr, Ca, Mg, Na, and Li while high $\omega_1$ values are observed for Cs, Rb, K, and Cl. This divergence coincides with the observation that ions in the low-frequency group tend to orient the dipole moment of water molecules in their first hydration shell, whereas ions in the high-frequency group do not.[124,125] Although this finding is new for *cations*, to the best of our knowledge, it mirrors a recent discovery, by correlated vibrational spectroscopy (CVS), of a divergence in the influence of anions on the frequency modes of water's H-bond network between 'hard' anions that strongly orient the dipole moment of water molecules in their first hydration shell and 'soft' anions that do not.[126]

Another significant trend that emerges from our observations is that several features of the dynamics of hydrated ions exhibit an ordering consistent with certain versions of the well-known Hofmeister series, a commonly-observed trend in the impact of ions in a wide range of phenomena including protein denaturation, the salting out of organic solutes, and the viscosity of electrolyte solutions.[1,68,125-127] For example, the stability of metal-ribonucleic acid (RNA) interactions follows a Hofmeister series of the form Cs $\lesssim$ Rb $\lesssim$ K $<$ Na $<$ Ba $<$ Li $<$ Sr $<$ Ca $<$ Mg.[1] This form matches both the divergence of $\omega_1$ values in Fig. 6a and the ordering of $\omega_2$ values in Fig. 6b (Na < Ba < Li ~ Sr < Ca). It also almost matches the ordering of $\omega_3$ values in Fig. 6c (Cs < Rb < K < Cl < Na < Mg < Ba $\lesssim$ Li $\lesssim$ Sr < Ca), albeit with a discrepancy on the placement of Mg. Finally, it matches the order of decreasing $\beta$ values in Table S3 (Cs ~ Rb ~ K ~ Cl $\gtrsim$ Na > Ba $\gtrsim$ Li ~ Sr > Ca > Mg). We note that slight variants of the Hofmeister series (differing, for example, in the relative placement of Li and Ba) have been reported for different properties. The consistency of our results with one of these variants suggest that at least some 'Hofmeister-like' effects, rather than reflecting a so-called 'structure making' impact of the ion, may instead reflect the coupled dynamics of the ion and its first hydration shell (and possibly its second hydration shell through the H-bond bend mode) in line with results by Heyden et al.[110] This finding parallels recent observations that at least some Hofmeister-like effects correlate with the tendency of ions to disrupt dipole-dipole correlations between water molecules up to their third hydration shell.[125]

Finally, as an additional conceptual perspective on the trends highlighted above, we define a 'dynamic hydration number' $N_\mathrm{w}^*$ such that the frequency $\omega_3$ exhibits an inverse-square-root mass dependence on the reduced mass $\mu_i^*$ of the ion and $N_\mathrm{w}^*$ water molecules (i.e., $\mu_i^* = \frac{m_i(N_\mathrm{w}^* m_\mathrm{w})}{m_i + N_\mathrm{w}^* m_\mathrm{w}}$). The resulting values of $N_\mathrm{w}^*$ have a relatively high statistical uncertainty (estimated at roughly $\pm$1/3 based on tests that used a narrower range of $m_i$ values in the fitting procedure). Nevertheless, they clearly exhibit a divergence between Cs, Rb, and K (for which $N_w^* > 8$) and all other cations (for which $N_w^* \approx 2 \pm 1$) as shown in Fig. S1. In other words, the high-frequency $\omega_3$ mode of Cs, Rb, and K exhibits features of a rattling motion decoupled from the dynamics of the hydration shell (as if their hydration shell were so massive that it did not not respond to the ion rattling motions), whereas for all other cations their high-frequency rattling mode has a mass-dependence conceptually consistent with the expected coupled dynamic of the ion and a rigid cluster of one to three water molecules. The behavior of Cl is intermediate between the two groups with $N_w^* \approx 4.2$.

## Acknowledgements

This work was supported by the U.S. Department of Energy, Office of Science, Office of Basic Energy Sciences (BES), Chemical Sciences, Geosciences, and Biosciences Division (CSGB), Geosciences Program under award DE-SC0018419. Computational resources were provided by Princeton University's shared Research Computing facilities.

## Disclaimer

## Author contributions

I.C.B. designed the study, performed the simulations, analyzed the data, wrote the original manuscript, and coordinated the research. M.-T.H.N. and T.R.U. contributed to the conceptual development and interpretation. C.V. contributed to the review and editing.

## Competing interests

There are no competing interests to declare.

**Table S1. Ion properties for all 10 simulated ions:** first maximum $r_{\max}$ and minimum $r_{\min}$ (Å) of the ion-water O radial distribution function; ion oxygen coordination number $N_O$; self-diffusion coefficient $D_i$ ($10^{-9}$ m$^2$ s$^{-1}$) of the major isotope; $\beta$ value expressing the mass-dependence of $D_i$; residence time of water in the first hydration shell $\tau_w$ (ps) of the major isotope; $\gamma$ value expressing the mass-dependence of $\tau_w$; mean-square force on the ion $\langle F^2 \rangle$ (nN$^2$). For Mg, the reported value of $\tau_w$ was calculated as the average over all five isotopes to decrease statistical uncertainty. The last two columns show values of $r_{\max}$ and $N_O$ determined in previous studies using the MADRID-2019 model.[46-48] Comparison with these results suggest that the precisions of $r_{\max}$ and $N_O$ values reported here are on the order of $\pm$0.01 and 0.1, respectively.

| | $r_{\max}$ | $r_{\min}$ | $N_O$ | $D_i$ | $\beta$ | $\tau_w$ | $\gamma$ | $\langle F^2 \rangle$ | $r_{\max}$* | $N_O$* |
|---|---|---|---|---|---|---|---|---|---|---|
| Li | 1.84 | 2.64 | 4.0 | 1.07(3) | 0.007(2) | 1,400(141) | 0.196(50) | 0.67 | 1.84 | 4.0 |
| Na | 2.33 | 3.13 | 5.5 | 1.36(4) | 0.029(13) | 16.6(9) | 0.047(27) | 0.41 | 2.33 | 5.5 |
| K | 2.74 | 3.54 | 6.8 | 1.93(3) | 0.033(8) | 6.4(1) | 0.052(14) | 0.29 | 2.73 | 6.5 |
| Rb | 2.74 | 3.54 | 6.4 | 1.97(5) | 0.039(7) | 5.8(1) | 0.047(18) | 0.23 | 2.75 | 6.5 |
| Cs | 2.86 | 3.70 | 6.9 | 2.03(4) | 0.037(9) | 5.1(1) | 0.052(11) | 0.20 | 2.87 | 6.6 |
| Mg | 1.92 | 3.19 | 6.0 | 0.83(4) | 0.001(2) | 75,500(5,900) | -0.026(89) | 1.07 | 1.92 | 6.0 |
| Ca | 2.39 | 3.21 | 7.4 | 0.89(3) | 0.003(2) | 102(7) | 0.022(27) | 1.21 | 2.38 | 7.4 |
| Sr | 2.60 | 3.40 | 8.1 | 0.85(4) | 0.007(3) | 130(11) | 0.027(25) | 1.02 | 2.60 | 8.0 |
| Ba | 2.84 | 3.68 | 9.0 | 0.87(5) | 0.009(8) | 60(2) | 0.035(9) | 0.82 | 2.84 | 9.0 |
| Cl | 3.03 | 3.66 | 5.9 | 1.69(3) | 0.038(4) | 8.0(1) | 0.050(2) | 0.31 | 3.04 | 5.9 |

**Table S2. Self-diffusion coefficient *$D_i$* ($10^{-9}$ m$^2$ s$^{-1}$) of the major isotope of each ion.** Results are corrected for finite-size effects as described in the text. Previous results obtained using the MADRID-2019 model illustrate the reproducibility of the results.[47] Experimental values are from the compilation reported by Li and Gregory.[102]

| | This work | MADRID-2019 | Experiments |
|---|---|---|---|
| $^{7}$Li | 1.07(3) | 1.07(9) | 1.03 |
| $^{23}$Na | 1.36(4) | 1.36(7) | 1.33 |
| $^{39}$K | 1.93(3) | 1.90(11) | 1.96 |
| $^{85}$Rb | 1.97(5) | 1.88(5) | 2.06 |
| $^{133}$Cs | 2.03(4) | 1.99(6) | 2.07 |
| $^{24}$Mg | 0.83(4) | 0.82(7) | 0.705 |
| $^{40}$Ca | 0.89(3) | 0.84(4) | 0.793 |
| $^{88}$Sr | 0.85(4) | | 0.794 |
| $^{138}$Ba | 0.87(5) | | 0.848 |
| $^{35}$Cl | 1.69(3) | 1.76(9) | 2.03 |

**Table S3. Values of the $\beta$ coefficient characterizing the mass dependence of *$D_i$*.** Results from previous studies were obtained using experiments (third column) or MD simulations (fourth column).

| | This work | Previous experiments | Previous simulations |
|---|---|---|---|
| Li | 0.007(2) | 0.015(2)[25], 0.0251(9)[104] | 0.017(15)[26] |
| Na | 0.029(13) | 0.023(23)[105] | 0.029(22)[27] |
| K | 0.033(8) | 0.042(2)[27], 0.0286(14)[104] | 0.049(17)[27] |
| Rb | 0.039(7) | 0.0287(9)[104] | |
| Cs | 0.037(9) | | 0.030(18)[27] |
| Mg | 0.001(2) | 0.0000(15)[25], 0.0015(1)[104] | 0.006(18)[26] |
| Ca | 0.003(2) | 0.0045(5)[27], 0.0064(2)[104] | 0.000(11)[27] |
| Sr | 0.007(3) | 0.0095(6)[104] | |
| Ba | 0.009(8) | 0.0129(8)[104] | |
| Cl | 0.038(4) | 0.026(14)[25], 0.030(3)[106] | 0.034(18)[27] |

**Table S4. Residence time $\tau_\mathrm{w}$ (ps) of water molecules in the first hydration shell of the major isotope of each ion.** Results are similar to those obtained in a previous study of seawater using the MADRID-2019 model, demonstrating minimal sensitivity to aqueous chemistry.[42] Results from other previous work are restricted to AIMD simulations or classical MD simulations carried out with the MB-nrg model (italics) and pump-probe spectroscopy measurements (regular font).

| | This work | MADRID-2019 seawater | Previous work (AIMD, pump-probe, or MB-nrg) |
|---|---|---|---|
| $^{7}$Li | 1,400(141) | | *>40*[107], *>100*[68] |
| $^{23}$Na | 16.6(9) | 18.6 | *~28*[68] |
| $^{39}$K | 6.4(1) | 6.7 | *~10*[107], *~7*[68] |
| $^{85}$Rb | 5.8(1) | | *9*[108]; *~6*[68] |
| $^{133}$Cs | 5.1(1) | | *7*[108]; *~5*[68] |
| $^{24}$Mg | 75,500(5,900) | 140,000 | *>60*[88] |
| $^{40}$Ca | 102(7) | 116 | *>60*[88], *~110*[90] |
| $^{88}$Sr | 130(11) | | |
| $^{138}$Ba | 60(2) | | |
| $^{35}$Cl | 8.0(1) | 7.7 | *12(3)*[98], 12(3)[109], ~6[100] |

**Table S5. Properties of all 50 isotopes of the 10 simulated ions:** ion mass $m_i$ (Da), ion self-diffusion coefficient obtained using the Green-Kubo or Einstein relations $D_{i,\mathrm{GK}}$, $D_{i,\mathrm{E}}$ ($10^{-9}$ m$^2$ s$^{-1}$); self-diffusion coefficient corrected for finite system size $D_i$ ($10^{-9}$ m$^2$ s$^{-1}$); residence time of water in the first hydration shell $\tau_\mathrm{w}$ (ps); Einstein frequency $\omega_\mathrm{E}$ (THz); frequencies of the three ion translational modes $\omega_1$, $\omega_2$, and $\omega_3$ (THz). As noted in the main text, for amplitude of the intermediate-frequency mode was consistently lower than that of the low and high-frequency modes. Therefore, where only two modes were unambiguously indentified, these are labeled $\omega_1$ and $\omega_3$. For $D_i$, $\tau_\mathrm{w}$, and $\omega_\mathrm{E}$, uncertainties are 95% confidence intervals based on the statistic error, evaluated by dividing each simulation into three 25-ns blocks. The most abundant isotope of each element is highlighted in bold font.

| | $m_i$ | $D_{i,\mathrm{GK}}$ | $D_{i,\mathrm{E}}$ | $D_i$ | $\tau_\mathrm{w}$ | $\omega_\mathrm{E}$ | $\omega_1$ | $\omega_2$ | $\omega_3$ |
|---|---|---|---|---|---|---|---|---|---|
| **$^{7}$Li** | **7.02** | **0.84(0)** | **0.82(1)** | **1.07(3)** | **1,400(141)** | **15.32(0)** | **1.09** | **5.73** | **15.04** |
| $^{14}$Li | 14.03 | 0.83(1) | 0.81(1) | 1.06(3) | 1,591(388) | 10.84(1) | 1.02 | 5.41 | 11.93 |
| $^{28}$Li | 28.06 | 0.82(2) | 0.81(2) | 1.06(4) | 1,859(44) | 7.67(1) | 0.96 | 4.63 | 10.19 |
| $^{56}$Li | 56.13 | 0.82(2) | 0.81(2) | 1.05(5) | 1,928(436) | 5.43(0) | 0.85 | 3.69 | 9.37 |
| $^{112}$Li | 112.26 | 0.82(1) | 0.80(0) | 1.05(3) | 2,506(639) | 3.84(0) | 0.69 | 2.95 | 9.09 |

| $^{6}$Na | 5.75 | 1.15(3) | 1.14(3) | 1.38(5) | 16.0(6) | 13.25(1) | 1.01 | 3.36 | 12.46 |
|---|---|---|---|---|---|---|---|---|---|
| $^{11}$Na | 11.49 | 1.16(3) | 1.14(3) | 1.39(5) | 15.4(5) | 9.36(0) | 0.99 | 3.27 | 9.68 |
| **$^{23}$Na** | **22.99** | **1.12(2)** | **1.13(1)** | **1.36(4)** | **16.6(9)** | **6.62(0)** | **0.93** | **2.71** | **7.61** |
| $^{46}$Na | 45.98 | 1.09(2) | 1.09(1) | 1.33(4) | 17.4(1.1) | 4.69(1) | 0.80 | 2.19 | 6.05 |
| $^{92}$Na | 91.96 | 1.05(2) | 1.03(3) | 1.28(4) | 17.7(9) | 3.32(1) | 0.72 | 1.69 | 5.80 |
| $^{5}$K | 4.87 | 1.83(3) | 1.84(3) | 2.06(5) | 5.8(5) | 12.03(1) | 1.66 | 5.85 | 11.04 |
| $^{10}$K | 9.74 | 1.81(3) | 1.82(1) | 2.04(5) | 5.9(3) | 8.51(0) | 1.59 | 5.94 | 8.98 |
| $^{19}$K | 19.48 | 1.77(1) | 1.77(3) | 2.00(3) | 6.0(4) | 6.02(0) | 1.43 | 4.82 | 7.03 |
| **$^{39}$K** | **38.96** | **1.70(1)** | **1.68(4)** | **1.93(3)** | **6.4(1)** | **4.25(0)** | **1.14** | | **4.32** |
| $^{78}$K | 77.93 | 1.66(3) | 1.67(3) | 1.89(5) | 6.7(4) | 3.02(0) | 0.79 | | 3.55 |
| $^{5}$Rb | 5.31 | 1.97(3) | 1.95(0) | 2.20(5) | 5.0(2) | 10.35(0) | 1.60 | 5.26 | 9.64 |
| $^{11}$Rb | 10.61 | 1.92(4) | 1.93(5) | 2.15(6) | 5.2(5) | 7.32(1) | 1.54 | 4.67 | 7.66 |
| $^{21}$Rb | 21.23 | 1.88(2) | 1.88(3) | 2.12(5) | 5.3(1) | 5.18(1) | 1.36 | 3.29 | 5.82 |
| $^{42}$Rb | 42.46 | 1.81(2) | 1.82(2) | 2.05(4) | 5.4(3) | 3.66(0) | 1.05 | | 3.27 |
| **$^{85}$Rb** | **84.91** | **1.74(2)** | **1.73(4)** | **1.97(5)** | **5.8(1)** | **2.60(0)** | **0.73** | | **3.28** |
| $^{8}$Cs | 8.31 | 2.02(2) | 2.04(4) | 2.25(4) | 4.5(2) | 7.80(0) | 1.48 | 4.62 | 7.74 |
| $^{17}$Cs | 16.61 | 1.99(8) | 1.99(7) | 2.22(11) | 4.5(2) | 5.52(0) | 1.38 | 3.99 | 4.79 |
| $^{33}$Cs | 33.23 | 1.94(2) | 1.95(3) | 2.17(5) | 4.7(2) | 3.90(0) | 1.11 | | 3.29 |
| $^{66}$Cs | 66.45 | 1.89(4) | 1.90(4) | 2.12(6) | 4.9(1) | 2.76(0) | 0.80 | | 2.74 |
| **$^{133}$Cs** | **132.91** | **1.80(2)** | **1.78(2)** | **2.03(4)** | **5.1(1)** | **1.96(0)** | **0.52** | | **1.78** |
| $^{6}$Mg | 6.00 | 0.60(1) | 0.59(2) | 0.83(3) | $71(109) \times 10^{3}$ | 21.00(1) | 1.06 | 5.09 | 15.79 |
| $^{12}$Mg | 11.99 | 0.60(1) | 0.59(0) | 0.83(3) | $85(123) \times 10^{3}$ | 14.85(0) | 1.02 | 6.07 | 12.07 |
| **$^{24}$Mg** | **23.99** | **0.60(1)** | **0.59(1)** | **0.83(4)** | **$81(111) \times 10^{3}$** | **10.50(0)** | **0.99** | **6.79** | **9.79** |
| $^{48}$Mg | 47.97 | 0.60(1) | 0.58(0) | 0.83(3) | $71(41) \times 10^{3}$ | 7.43(0) | 0.91 | 5.01 | 7.57 |
| $^{96}$Mg | 95.94 | 0.59(1) | 0.59(2) | 0.83(3) | $71(109) \times 10^{3}$ | 5.27(0) | 0.77 | 4.36 | 6.44 |
| $^{5}$Ca | 5.00 | 0.67(2) | 0.66(2) | 0.90(4) | 92(4) | 24.45(2) | 0.99 | 6.65 | |
| $^{10}$Ca | 9.99 | 0.66(1) | 0.65(2) | 0.90(3) | 94(2) | 17.31(2) | 0.98 | 6.49 | 17.71 |
| $^{20}$Ca | 19.98 | 0.66(1) | 0.65(1) | 0.89(3) | 96(5) | 12.25(1) | 0.94 | 6.02 | 13.78 |
| **$^{40}$Ca** | **39.96** | **0.66(0)** | **0.65(1)** | **0.89(3)** | **102(7)** | **8.66(1)** | **0.86** | **5.31** | **11.68** |
| $^{80}$Ca | 79.93 | 0.66(1) | 0.65(0) | 0.89(3) | 96(2) | 6.13(1) | 0.76 | 4.16 | 10.70 |
| $^{5}$Sr | 5.49 | 0.63(0) | 0.63(1) | 0.86(3) | 118(1) | 21.43(0) | 0.97 | 5.11 | 20.72 |
| $^{11}$Sr | 10.99 | 0.63(1) | 0.62(1) | 0.86(3) | 124(4) | 15.17(0) | 0.91 | 5.04 | 15.50 |
| $^{22}$Sr | 21.98 | 0.63(1) | 0.62(1) | 0.86(4) | 119(1) | 10.72(0) | 0.88 | 4.64 | 12.00 |
| $^{44}$Sr | 43.95 | 0.62(0) | 0.62(0) | 0.85(3) | 123(3) | 7.59(0) | 0.83 | 4.01 | 10.00 |
| **$^{88}$Sr** | **87.91** | **0.62(1)** | **0.61(2)** | **0.85(4)** | **130(11)** | **5.38(0)** | **0.75** | **3.23** | **8.95** |
| $^{9}$Ba | 8.62 | 0.65(2) | 0.65(2) | 0.88(4) | 53.7(8) | 15.35(0) | 0.86 | 3.76 | 15.27 |
| $^{17}$Ba | 17.24 | 0.66(1) | 0.65(1) | 0.89(3) | 55.1(4) | 10.85(0) | 0.82 | 3.59 | 11.60 |
| $^{34}$Ba | 34.48 | 0.65(1) | 0.64(1) | 0.89(3) | 55.6(2.0) | 7.68(0) | 0.79 | 3.09 | 9.28 |
| $^{69}$Ba | 68.95 | 0.64(1) | 0.63(0) | 0.88(3) | 57.2(1.8) | 5.44(0) | 0.75 | 2.57 | 7.78 |
| **$^{138}$Ba** | **137.91** | **0.63(3)** | **0.63(3)** | **0.87(5)** | **59.6(1.9)** | **3.85(1)** | **0.66** | **2.01** | **7.01** |
| $^{9}$Cl | 8.74 | 1.53(2) | 1.52(3) | 1.77(4) | 7.5(1) | 9.40(0) | 1.75 | 6.66 | 9.06 |
| $^{17}$Cl | 17.48 | 1.49(4) | 1.49(5) | 1.72(7) | 7.8(1) | 6.65(0) | 1.56 | 5.11 | 7.12 |
| **$^{35}$Cl** | **34.97** | **1.46(0)** | **1.45(1)** | **1.69(3)** | **8.0(1)** | **4.71(0)** | **1.26** | | **5.08** |
| $^{70}$Cl | 69.94 | 1.40(2) | 1.40(3) | 1.63(5) | 8.3(1) | 3.33(0) | 0.90 | | 4.32 |
| $^{140}$Cl | 139.88 | 1.35(1) | 1.36(4) | 1.59(3) | 8.6(1) | 2.36(0) | 0.60 | | 4.19 |

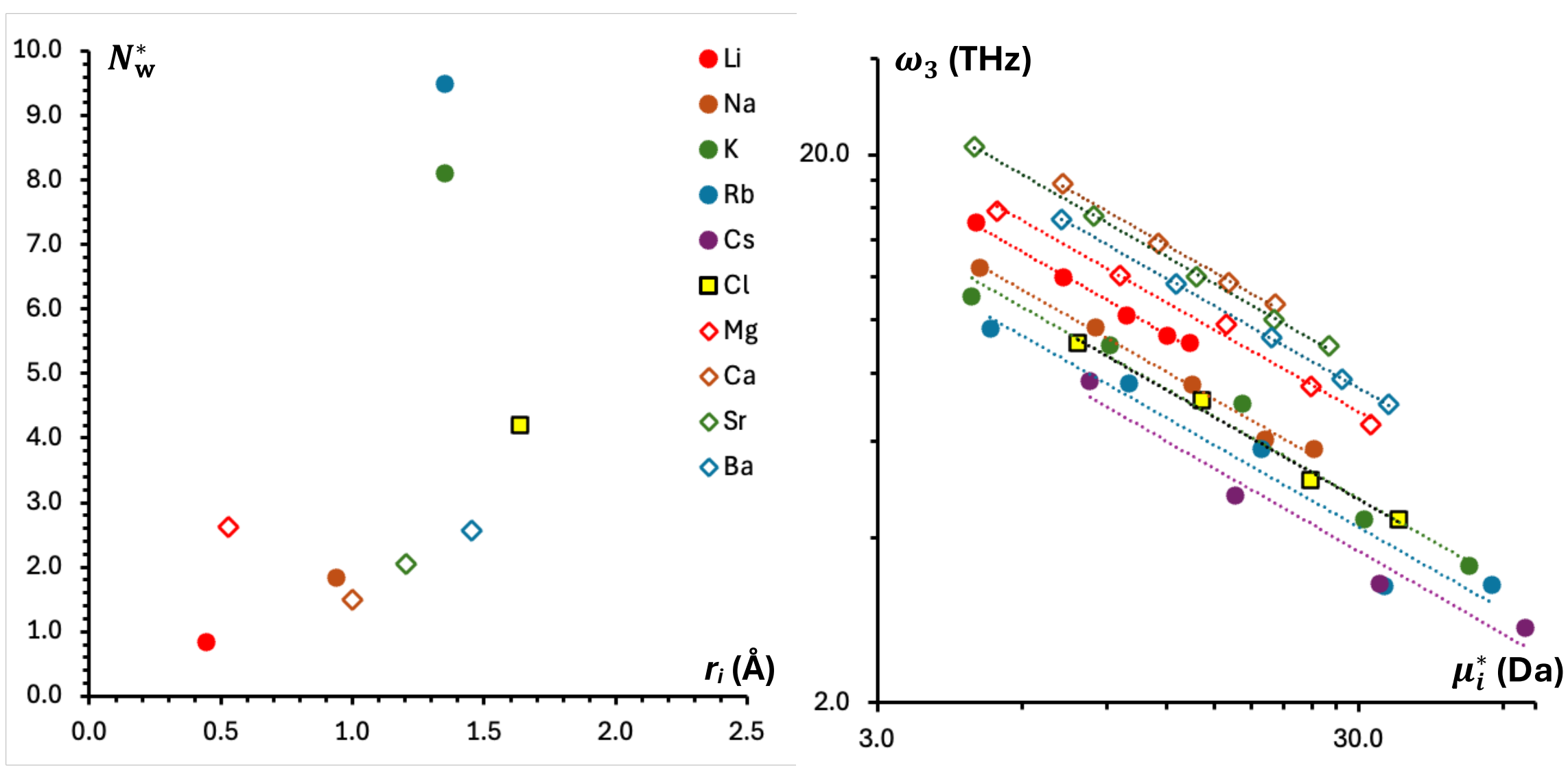

**Figure S1. Evaluation of the dynamic hydration number $N_w^*$ based on frequency $\omega_3$ (THz). (a)** Dynamic hydration number $N_w^*$ as a function of bare ion radius $r_i$ (Å). Cs is off-scale with $N_w^* > 10$. As noted in the main text, the statistical uncertainty of calculated $N_w^*$ values is estimated as roughly ±1/3. **(b)** Frequency $\omega_3$ as a function of the reduced mass $\mu_i^*$ (i.e., the reduced mass of the ion and a cluster of $N_w^*$ water molecules) plotted on a double logarithmic scale. The value of $N_w^*$ was determined such that a linear regression of $\log \omega_3$ vs. $\log \mu_i^*$ yields a slope of -0.5. Values are not included for $^{5}$Ca, for which the fourth mode at ≈ 28 THz prevents precise quantification of $\omega_3$, or for the two heaviest monovalent ions ($^{133}$Cs, $^{140}$Cl), for which the amplitude of the $\omega_3$ mode is nearly zero and its frequency is poorly defined.